# Magneto-elastic effect in impact dynamics of non–Newtonian ferrofluid droplets

**Gudlavalleti V V S Vara Prasad [a], Purbarun Dhar [b,*, 2] and Devranjan Samanta [a,*, 1]**

[a] Department of Mechanical Engineering, Indian Institute of Technology Ropar,

Rupnagar–140001, Punjab, India

[b] Department of Mechanical Engineering, Indian Institute of Technology Kharagpur,

Kharagpur–721302, West Bengal, India

*Corresponding authors:

[1]E-mail: devranjan.samanta@iitrpr.ac.in

[1]Tel: +91-1881-24-2109

[2]E-mail: purbarun@mech.iitkgp.ac.in; purbarun.iit@gmail.com

[1]Tel: +91-3222-28-2938

## Abstract

In this article, we propose, with the aid of detailed experiments and scaling analysis, the existence of magneto-elastic effects in the impact hydrodynamics of non-Newtonian ferrofluid droplets on superhydrophobic (SH) surfaces in presence of a magnetic field. The effects of magnetic Bond number ($Bo_m$), Weber number ($We$), polymer concentration and magnetic nanoparticle ($Fe_3O_4$) concentration in the ferrofluids were investigated. In comparison to Newtonian ferrofluid droplets, addition of polymers caused rebound suppression of the droplets relatively at lower $Bo_m$ for a fixed magnetic nanoparticle concentration and $We$. We further observed that for a fixed polymer concentration and $We$, increasing magnetic nanoparticle concentration also triggers earlier rebound suppression with increasing $Bo_m$. In the absence of the magnetic nanoparticles, the non-Newtonian droplets do not show rebound suppression for the range of $Bo_m$ investigated. Likewise, the Newtonian ferrofluids show rebound suppression at large $Bo_m$. This intriguing interplay of elastic effects of polymer chains and the magnetic nanoparticles, dubbed as the *magneto-elastic effect* is noted to lead to the rebound suppression. We establish a scaling relationship to show  that the rebound suppression is observed as manifestation of onset of magneto-elastic instability only when the proposed magnetic Weissenberg number ($Wi_m$) exceeds unity. We also put forward a phase map to identify the various regimes of impact ferrohydrodynamics of such droplets, and the occurrence of the magneto-elastic effect.

*Keywords:* non-Newtonian fluids; ferrofluid; ferrohydrodynamics; superhydrophobicity; rebound suppression

## 1. Introduction

The dynamics and phenomenology of droplet impact on a solid or liquid surface is scientifically important due to the wide range of associated applications. Droplet impact dynamics comprises various physical phenomena, like deposition, rebound, partial rebound [1], splashing and fragmentation. Understanding the dynamics is a decisive factor in many applications like inkjet printing, spray coating and painting, spray cooling, and retention of pesticide sprays on vegetation to prevent groundwater contamination and pesticide over-use, etc. The manipulation of impact dynamics of ferrofluid droplets by magnetic field can lead to significant improvements in magnetic 3D printing [2] performance, and may also promote control of shape distortions in metallic droplets during welding or soldering. The magnetic force on ferrofluid droplets is important for handling magnetic beads and manipulating ferrofluid droplet transport and splitting in microfluidic devices [3]. Ahmed et al. [4] studied the maximum spreading dynamics of a ferrofluid droplet under the effect of vertically oriented magnetic field. The maximum spreading of the droplet was proportional to the corresponding non-dimensional numbers, like Weber number ($We$), magnetic Bond number ($Bo_m$), and Reynolds number ($Re$).

Sudo et al. [5] explored the effect of magnetic field on the maximum spreading diameter and spike formation within the liquid lamella of impacting magnetic fluid droplets. Rahimi & Weihs [6] reported the droplet impact dynamics of magnetorheological fluids (MRF) and reported the dependence of maximum spreading factor on the magnetic field strength and Reynolds number ($Re$). Zhou and Jing [7] showed how magnetic field affects the collision characteristics, oscillation kinetics, maximum spreading factor, maximum recoiling height and the retraction height of ferrofluid droplets for different impact velocities. Sahoo et al. [8] experimentally investigated impact dynamics of ferrofluid droplets on superhydrophobic surfaces under the influence of horizontal magnetic field and reported that significant rebound suppression phenomena was observed at moderate magnetic Bond numbers ($Bo_m$ ~300). The ferrofluid droplet liquid lamella was shown to become largely unstable due to nucleation of holes during retraction stage at higher magnetic Bond numbers ($Bo_m$ >300), leading to distinct fragmentation kinetics fairly uncommon in droplet impact literature.

In recent years, elastic and viscoelastic effects during impact dynamics of non-Newtonian fluid droplets have gathered attention due to the observation of rebound suppression on superhydrophobic surfaces [9–12]. It was shown that addition of minute amounts of flexible, very long-chain polymers like polyethylene oxide (PEO) or polyacrylamide (PAM) to water arrests the droplet rebound on superhydrophobic surfaces. Initially it was believed to be due to the higher energy dissipation caused by the elongational viscosity of the polymer solutions. Bartolo et al. [9] proposed that during retraction, the polymer chains generate a large amount of normal stresses, and slows down the moving contact line of the droplet. This opposes the capillary force and retraction kinetic energy, which ultimately leads to rebound suppression. Mao et al. [10] showed the dependence of rebound behavior on the spread factor and the static contact angle of the droplet. Smith et al. [11] showed the extensional deformation of solvated fluorescent dyed DNA molecules near the receding contact line of a droplet slows down the retraction process.

Later, Dhar et al. [12] showed the governing roles of the impact velocity and polymer concentrations as critical parameters to determine the onset of rebound suppression. Based on the shear rate during onset of retraction and the relaxation time scale of the elastic fluid, they showed that the onset of rebound suppression occurs only under circumstances where the governing Weissenberg number (Wi) exceeds one. They further showed that the change in retraction dynamics of the non-Newtonian droplets can be also used to morph the thermo-species transport behavior, such as delaying the droplet Leidenfrost effect [13]. Zang et al. [14] observed a transition from droplet rebound to sticking by the addition of PEO to Newtonian fluids and reported the importance of sliding angle of the fluid droplet, in addition to the contact angle. Yun et al. [15] reported the effect of electric field on non-axisymmetric droplets towards inhibiting droplet rebound and also investigated the impact dynamics of ellipsoidal drops as function of the geometric aspect ratio and impact Weber number [16]. Antonini et al. [17] showed the importance of the receding contact angle of the rebounding droplet on hydrophobic and superhydrophobic surfaces. Very recently a work revisited the role of elongational viscosity on rebound suppression event using advanced experimental techniques like total internal reflection microscopy [TIRM] [18].

In this article, we explore a novel phenomenon in non-Newtonian ferrofluid droplets, which may be given a nomenclature of magneto-elastic effect. We investigate the impact and rebound dynamics of polymer solution based ferrofluid droplets under the influence of a horizontal magnetic field, and the phenomenology of onset of conjugated magnetic and elastic instability, leading to suppressed droplet rebound. The impact magneto-hydrodynamics of the droplet has been characterized by four dimensionless parameters, viz. the Weber number ($We = \rho V_0^2 D_0/\sigma_{lv}$, defined as the ratio of inertial force to surface tension force), the magnetic Bond number ($Bo_m = B^2 D_0/\mu_0 \sigma_{lv}$, defined as the ratio of magnetic force to surface tension force), the magnetic capillary number ($Ca_m = \eta V_0/\sigma_{lv}$, is defined as ratio of viscous force to surface tension forces under influence of magnetic field), and the Hartmann number ($Ha = \rho m B D_0/2V_0 \eta$, [19], defined as the ratio of magnetic force to viscous force). Here $\rho, V_0, D_0, \sigma_{lv}, B, m, \mu_0,$ and $\eta$ denote the density, impact velocity, pre-impact diameter, surface tension, magnetic flux density, magnetic moment, magnetic permeability of free space and viscosity of the ferrofluid droplet, respectively. Furthermore, we theorize and show that the onset of suppression of droplet rebound depends on the magneto-elastic effect, which is a conjugation of the polymer relaxation time, the shear rate at the contact line during the retraction stage, and the ferro hydrodynamics of the droplet. We also highlight the critical role of a proposed magnetic Weissenberg number in governing the onset criterion for droplet rebound suppression.

## 2. Materials and Methodologies

### 2.1. Non-Newtonian ferrofluids

The ferrofluids were synthesized using iron (II, III) oxide ($Fe_3O_4$) nanoparticles (Alfa Aesar India, >98.5% purity) of average particle diameter 20–30nm (noted from scanning electron microscopy images (not illustrated)). Polyethylene glycol (PEG 400, analytical grade, Finar Chemicals, India) was used to render the ferrofluids non-Newtonian. Initially, polymer solutions using PEG and deionized (DI) water were synthesized (5 and 10 % v/v of PEG). To these non-Newtonian fluids, the $Fe_3O_4$ nanoparticles are dispersed (2.5 and 5 wt. %) and stabilized following similarly reported protocol [20]. The polymeric ferrofluids were subjected to ultra-sonication for 2 hours to ensure colloidal stability. We adopt a nomenclature method for the polymeric ferrofluids: a sample containing '$x$' % PEG in water, and '$y$' % $Fe_3O_4$ particles is written as Px–Fy. Thus, sample P5-F2.5 contains 5 % v/v PEG in water with 2.5 w/w % particles dispersed in it. Any sample with P0 signifies Newtonian ferrofluid.

### 2.2. Substrates

Glass slides coated with a commercial superhydrophobic (SH) coating (Neverwet Ultra Ever dry, USA) were used as substrates in the experimental study. The substrates were prepared by following previously reported protocol [21]. Before spray coating, each glass slide was cleaned with DI water followed by acetone, and oven-dried thoroughly. The static equilibrium contact angles of the ferrofluid droplets are shown in Fig. S1 (refer supporting information). The static contact angle of non-Newtonian ferrofluid droplets on the SH surface is in the range of $155 \pm 3^0$. The surface tension of the ferrofluids (refer supporting information [Table. S1]) was measured using the pendant drop method, and the equilibrium static contact angles were measured from image analysis. The surface tension values were noted to be minimally altered due to the magnetic field, and in the range of $74 \pm 3$ mN/m.

### 2.3. Experimental setup

The overall arrangement of the experimental setup is shown in Fig. 1. A digitized droplet dispensing mechanism (DDM) unit (± 0.1μl volumetric accuracy) was used to dispense droplets of fixed volume, via a glass micro-syringe with a flat head steel needle (22 gauge). The impact height of the droplet was adjusted to obtain different impact velocities, and different *We*. The droplet was allowed to fall freely on the substrate, which was placed between the poles of an electromagnet. An electromagnet (Holmarc Opto-Mechatronics Ltd., India) was used to generate the magnetic field, with controlled field strengths of 0–0.2 T. The field strength was varied by changing current input across the electromagnet pole windings using a direct current power supply (Polytonic Corp., India) unit. Flat face cylindrical iron billets of diameter 100 mm act as the magnetic pole shoes. The magnetic field strength near the substrate was calibrated by a Gauss meter (Holmarc Opto-Mechatronics Ltd., India) for different current inputs to the pole windings. Beyond 0.2 T, the free falling droplet shape is distorted by the field before impact, which induces artefacts to the post impact hydrodynamics. Accordingly, the experiments have been restricted to 0.2 T.

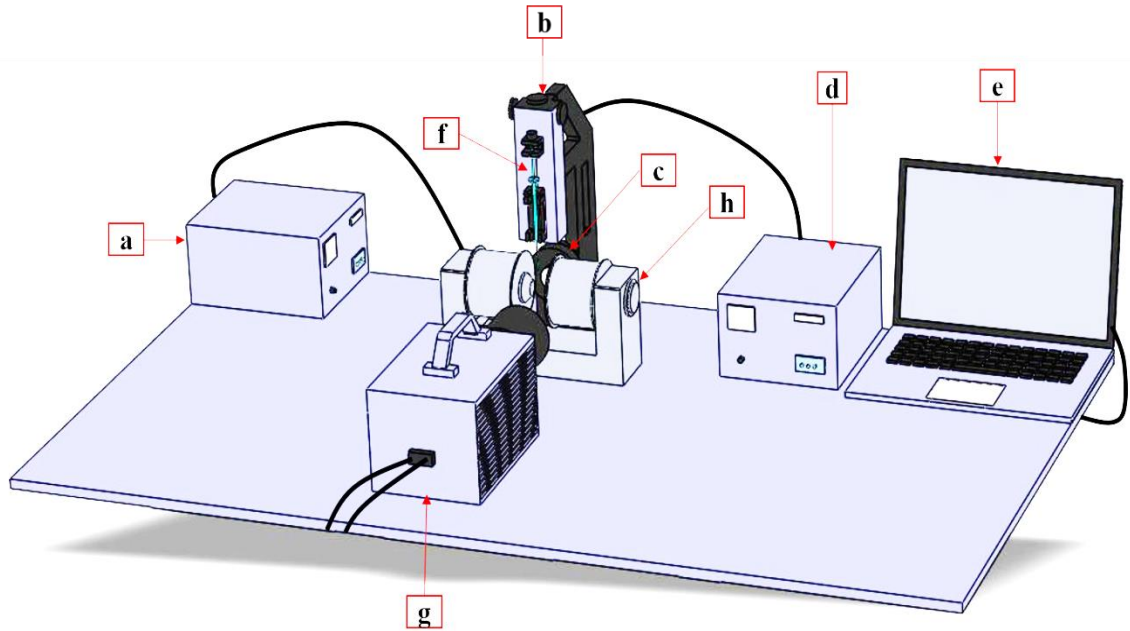

**Fig. 1:** Schematic of the experimental setup. (a) Electromagnet power controller (b) droplet dispensing mechanism (DDM) unit (c) LED array for backlight illumination (d) DDM and LED array intensity controller (e) computer for data acquisition and camera control (f) micro-syringe (g) high speed camera (h) programmable electromagnet unit

The micro-syringe needle and SH slides were positioned at the center of the electromagnet poles to ensure that the droplet impact was in a uniform magnetic field. The schematic of the front and top views of positioning of the droplet, and the associated coordinate frame of reference are illustrated in Fig. S2 (supporting information). The droplet dynamics was recorded with a high-speed camera (Fastcam SA4, Photron, UK) attached with a macro lens of constant focal length of 105 mm (Nikkor, Nikon). All experiments are conducted at 3600 frames per second (fps) and 1024 x 1024 pixels resolution. For backlight illumination, an intensity controlled LED array was used. Rheological tests of the fluids were performed using a rotational rheometer (Anton Paar, Germany). A plate and plate geometry (PP-20) attached to a magnetorheological module is used to determine the shear viscosity ($\eta$), elastic modulus ($G^{'}$) and viscous modulus ($G^{''}$) (refer Fig. S3 and S4 in supporting information) under influence of magnetic field. All tests and experiments were repeated thrice to check for repeatability.

## 3. Results and discussions

### *3.1. Droplet impact ferrohydrodynamics*

We start the discussions with the ferrohydrodynamics of the droplet impact event on SH surfaces. Fig. 2 (a), (b) and (c) show the effect of the magnetic Bond number on the ferrofluid droplet spreading due to increasing non-Newtonian nature (achieved by increasing polymer concentration) at *We* ~100. In both Newtonian (water based) and non-Newtonian (polymer solution based) ferrofluid droplet cases, the droplet spreads with radial symmetry intact in the absence of magnetic field ($Bo_m$ =0). For the Newtonian case, the droplets spread to a larger extent along the z direction compared to the x direction (refer fig. 2 c, 1st row, 5th column for the coordinate axis) with increasing magnetic Bond number ($Bo_m$). This is due to the interplay between magnetic forces and the surface tension force, and has been discussed in details in our previous report [8]. As the droplet spreads in presence of the magnetic field, the rate of change of the magnetic force on the droplet increases. The Lorentz force, which suppresses the cause of change in the magnetic state of the system, thereby opposes the spread in the x-direction, while the droplet spreads unrestrictedly along the z-direction (refer fig. 2 for x and z directions). Additionally, the liquid film formed and enclosed by the spreading rim (fig. 2a, 4th row, 3rd column) undergoes rupture due to nucleation and proliferation of holes at high values of magnetic Bond number (at $Bo_m$ ~1200) [8]. We discuss the genesis of the rupturing instability from energy conservation considerations in the previous report [8]. To quantify the asymmetric spreading kinetics, we define the $\xi_{max}$ (= $(D_z/D_x)_{max}$, the maximum non-dimensional orthogonal spreading) and illustrate the same in fig. 3.

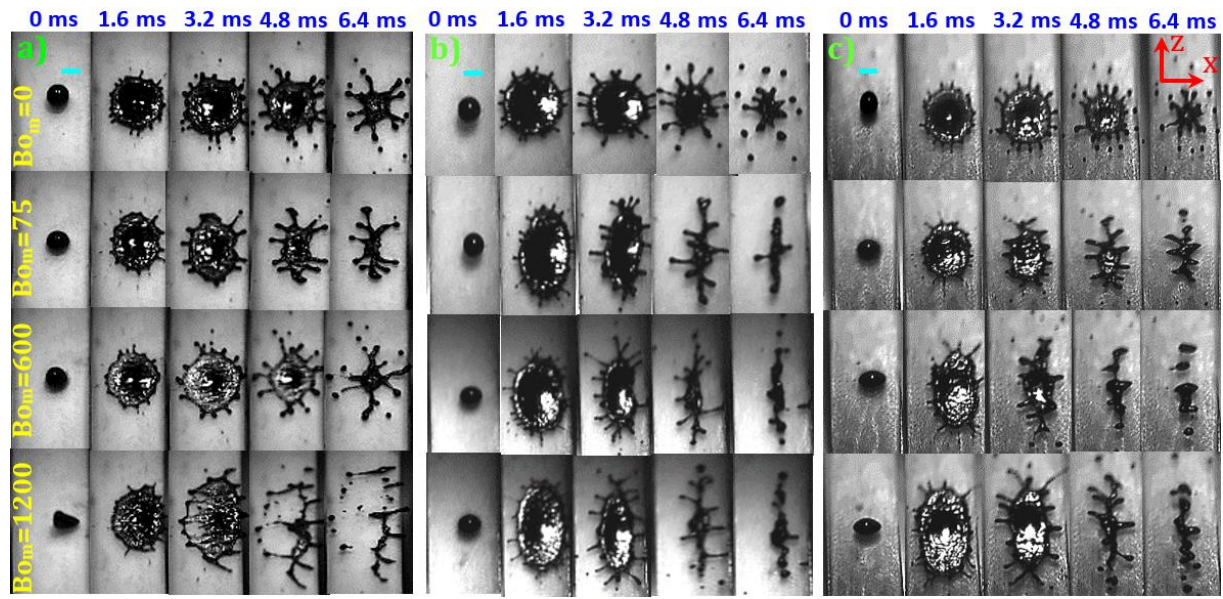

**Fig. 2:** Temporal evolution of droplet impact ferrohydrodynamics for different $Bo_m$ for (a) P0-F2.5 (b) P5-F2.5 and (c) P10-F2.5 fluids. The figure shows the role of increasing non-Newtonian fluid nature on the droplet impact ferrohydrodynamics. The scale bars represent

2.8 mm. The images are 1.6 ms apart. The magnetic field acts along the x-direction (coordinate axis shown on top right)

Now we shift the focus on the non-Newtonian ferrofluids. The impact phenomenon is characterized by the formation of distinct filaments. This can be observed from fig 2b and c. With increase in either polymer concentration or the $Bo_m$, the rupturing instability of the spread droplet is noted to be seized. Also, the asymmetric elongation along the z-direction is notably pronounced. Figure 3 illustrates the non-dimensional orthogonal maximum spreading for both Newtonian and non-Newtonian ferrofluid droplets for varied magnetic field intensity, such as 0 T ($Bo_m$ =0), 0.05 T ($Bo_m$ ~75), 0.10 T ($Bo_m$ ~600) and 0.20 T ($Bo_m$ ~1200). As the physical properties of the fluids vary to some extent, the $Bo_m$ are not exactly same, and hence we use the magnetic field intensity to showcase the data. For a fixed magnetic particle (F2.5) concentration, the $\xi_{max}$ increases with increasing polymer concentration (fig. 3a). At the same time, for a fixed polymer (P10) concentration, $\xi_{max}$ increases with increasing magnetic particle concentration (fig. 3b). In a previous report [22] it has been shown that PEG chains may entangle $Fe_3O_4$ nanoparticles to form a particle–polymer chain mesh. We believe, in the non-Newtonian ferrofluids, the particles entangled to the fluid phase are able to arrest the spread along the x-direction to a greater extent. This is possible as the Lorentz force acts on the whole fluid in a more uniform manner compared to the Newtonian case where the particles are dispersed, and prone to magnetophoretic drift with respect to the fluid. Consequently, the droplet spreads to a larger extent along the z-direction, thereby increasing the $\xi_{max}$.

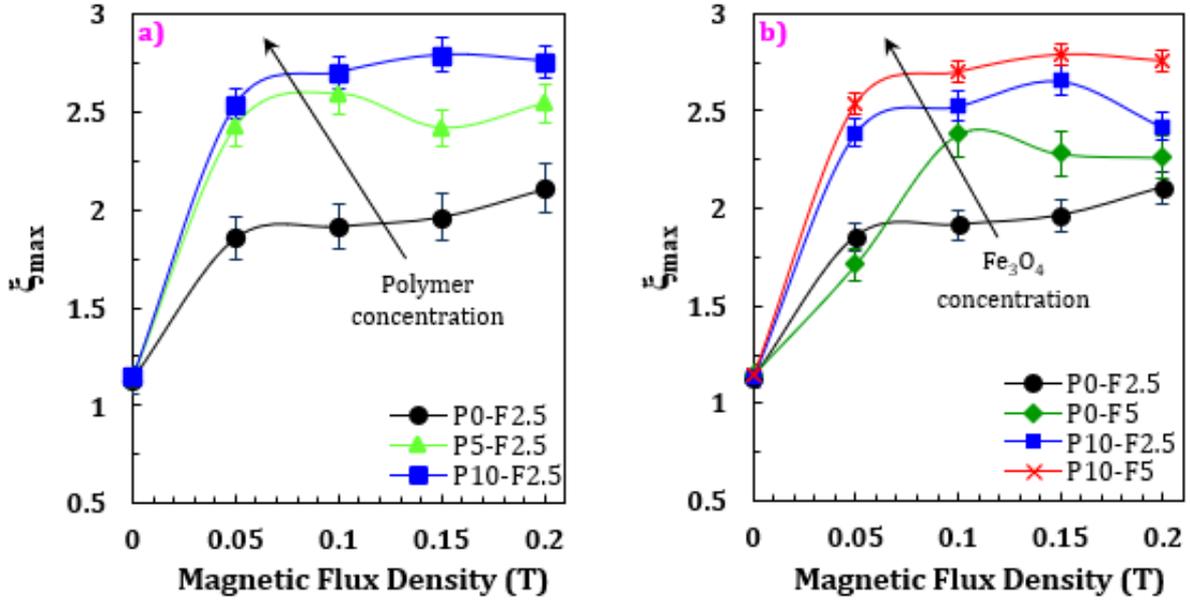

**Fig. 3:** The effect of magnetic flux density (T) on the maximum non-dimensional orthogonal spreading ($\xi_{max}$) with (a) varying polymer concentration and (b) magnetic particle concentration.

## 3.2. Droplet rebound suppression kinetics

In this section, we shall discuss the phenomenology and physical mechanisms responsible for the suppression of rebound of the droplets under the effect of magnetic field.

### 3.2.1. Role of polymer concentration

The rebound phenomenology of the ferrofluid droplets in presence of horizontal uniform magnetic field on SH surfaces have been shown in Fig. 4. The set of experiments reported were done at fixed Weber number $We$ ~100 and magnetic nanoparticle concentration (F2.5), and the figure illustrates the role of polymer concentration (non-Newtonian behavior). Additionally, as a set of control experiments (not illustrated), we perform the impact studies for only polymer solutions (P5-F0 and P10-F0) for different $We$ and $Bo_m$, and no rebound suppression is noted at all. It can be readily observed that in absence of magnetic field ($Bo_m$ =0), both Newtonian and non-Newtonian ferrofluid droplets exhibit the usual droplet rebound behavior (fig. 4a, b and c, 1st row). In case of Newtonian ferrofluid, increasing the $Bo_m$ reveals the following sequence of droplet impact outcomes, such as spreading, retraction, recoil, rebound and fragmentation. Under no circumstances the non-Newtonian ferrofluid exhibits rebound suppression for the studied range of $Bo_m$.

Next, we focus on the non-Newtonian ferrofluids, and keep the magnetic nanoparticle concentration fixed (F2.5) while varying the polymer concentration. The lower polymer concentration (P5) ferrofluid droplets also show similar impact phenomenology as the Newtonian ferrofluid droplets up to $Bo_m$ ~600 (fig. 4b). But, at higher magnetic field strength ($Bo_m$ ~1200), onset of rebound suppression of the droplet was observed (fig. 4b, 5th row). In the case of higher polymer concentrations (P10), post-impact stages similar to P0 and P5 ferrofluids are noted in absence of magnetic field. But interestingly, the P10 ferrofluids exhibit the onset of rebound suppression from lower magnetic field strength regime ($Bo_m$ ~300) (fig 4c, 3rd row). For ease of illustration, we have enclosed the paradigms of rebound suppression in fig. 4 with dashed lines. Thereby, we infer that at fixed impact $We$ and magnetic particle concentration, the presence of certain non-Newtonian effect in conjunction with the magnetic force on the ferrofluid, triggers the rebound suppression phenomenon.

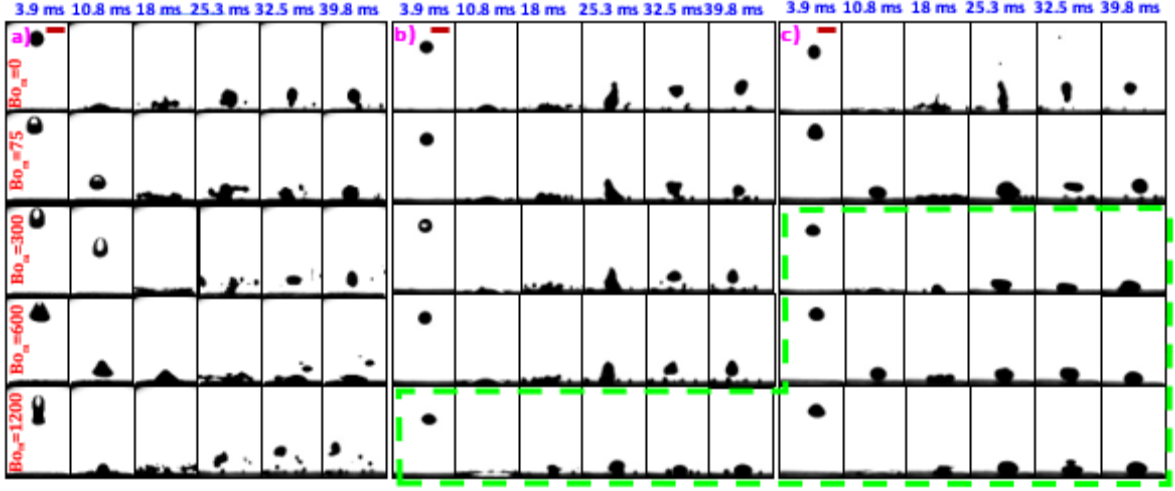

**Fig. 4:** Effects of polymer concentration on the rebound kinetics of the ferrofluid droplets over SH surface for different magnetic Bond numbers at *We* ~100 for (a) P0–F2.5 (b) P5–F2.5 and (c) P10–F2.5. The scale bars represent 2.8 mm. From top to bottom, the rows represent $Bo_m$=0, 75, 300, 600 and 1200. The region circumscribed by the dashed lines illustrates the regimes of rebound suppression.

### *3.2.2. Role of magnetic particle concentration*

Next, we probe the role of the magnetic particle concentration, which governs the magnetic force on the ferrofluid at a particular magnetic field (illustrated in fig. 5). We fix the polymer concentration (P10) and the Weber number (*We* ~100), and vary the magnetic particle concentration (F2.5 and F5). In the absence of magnetic field, both the ferrofluid droplets (P10-F2.5 and P10-F5) show the typical rebound nature off SH surfaces (fig. 5a and b, 1st row). The low concentration (F2.5) non-Newtonian ferrofluid droplet shows onset of rebound suppression at moderate $Bo_m$ ~300. But in case of a high concentration ferrofluid (F5) the same is noted at $Bo_m$ ~75. Our control experiments using Newtonian ferrofluid droplets of F2.5 and F5 do not show any such rebound suppression events, even at $Bo_m$~1200. Hence from the observations, we may further infer that the rebound suppression event is triggered by interplay of the magnetic and the non-Newtonian effects in the presence of a magnetic field.

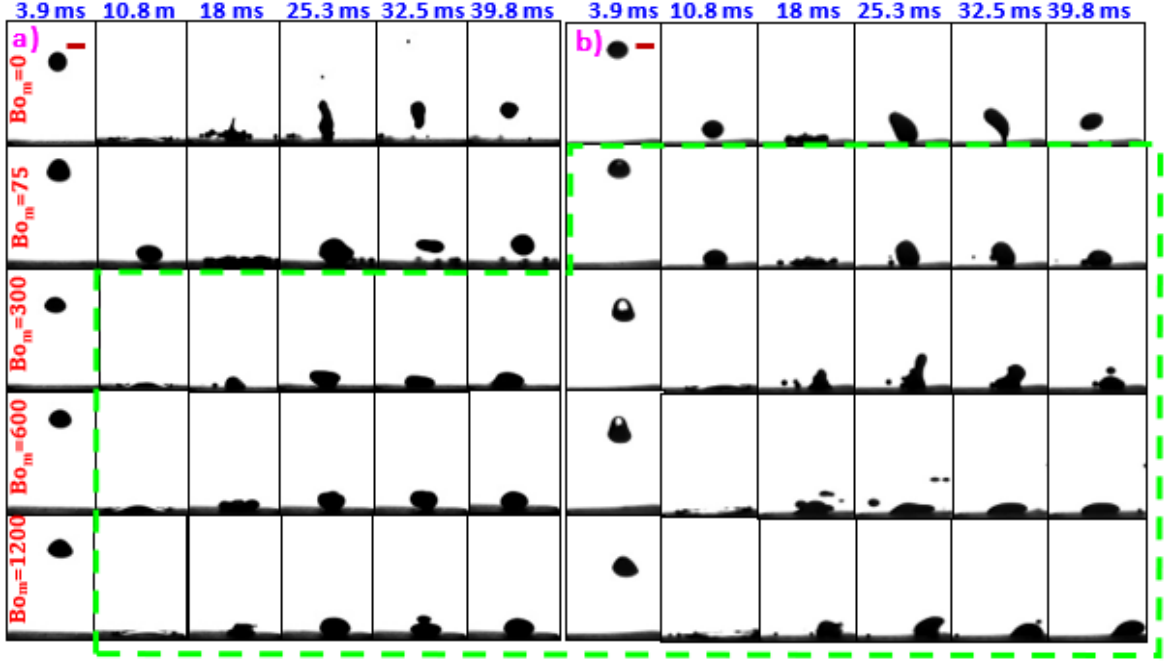

**Fig. 5:** Effects of magnetic particle concentration on the rebound suppression over SH surface for different magnetic Bond numbers, for (a) P10-F2.5 and (b) P10-F5, at *We* ~100. The scale bar represents 2.81 mm. From top to bottom, the rows represent $Bo_m$=0, 75, 300, 600 and 1200. The region circumscribed by the dashed lines illustrates the regimes of rebound suppression.

*3.2.3. Role of impact Weber number*

In this section, we discuss the role of the impact *We*. The impact phenomenology of different non-Newtonian ferrofluid droplets at two different *We*~40 and ~100 have been illustrated in fig. 6. Our experiments show (not illustrated) that when the polymer concentration is kept constant and magnetic particle concentration is varied, the onset of rebound suppression under magnetic field influence is not governed by the *We*. But the same is not true for the case where magnetic particle concentration is kept fixed, and the polymer concentration is varied (fig. 6). At a fixed We~40, irrespective of the nature of the ferrofluid (Newtonian or non-Newtonian) and the $Bo_m$, the droplets do not exhibit any sign of rebound suppression. When the same set of impact experiments are conducted at We~100, we note significant and drastic change in the associated hydrodynamics. At higher *We*, the onset of rebound suppression is observed at $Bo_m$ ~600 for low polymer concentration (P5), and at $Bo_m$ ~75 for high polymer concentration (P10) cases. We therefore infer that the impact *We* and the non-Newtonian characteristics of the fluid (polymer concentration) also interplay, and higher *We* lead to triggering of the magneto-elastic effect. This observation is in agreement to our previous report on the elastic instability in non-Newtonian droplets [12].

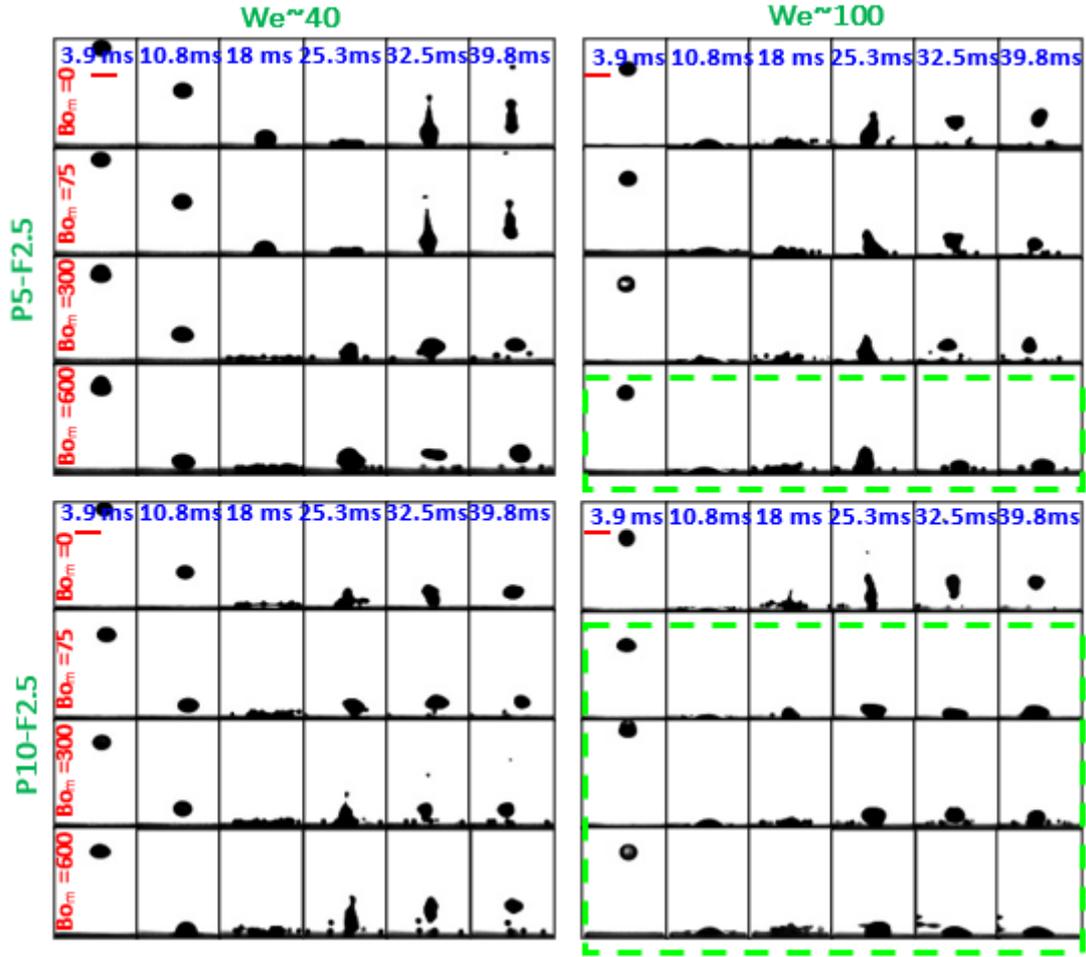

**Fig. 6:** Effects of the impact Weber number on the impact events over SH surface for different $Bo_m$ at We~40 and ~100. The scale bar represents 2.81 mm. From top to bottom, the rows represent $Bo_m$=0, 75, 300, and 600. The regions circumscribed by the dashed lines illustrate the regimes of rebound suppression.

*3.2.4. The 'magneto-elastic' effect and role of magnetic Weissenberg number*

From the discussions in the previous three subsections, we infer that an interplay of the impact mechanics (manifested through the *We*), the elasticity of the fluid (manifested through the polymer concentration), and the ferrohydrodynamic forces (manifested through the $Bo_m$) are responsible for the triggering of the rebound suppression event. It is at this instance that the plausible role of viscosity of the fluids on the rebound suppression dynamics needs to be discussed. In our previous reports [12, 13] on elastic fluids, we have shown conclusively that the increase in viscosity of the fluids due to presence of polymer may alter the impact hydrodynamics, but does not trigger rebound suppression under any circumstances. In the present study, we have performed detailed rheology and magneto-rheology (fig. S2, supporting information) to understand the role played by viscosity. As a representative case, we shall discuss the case for fluids P5-F2.5 at 0.1 T (corresponding to $Bo_m$ ~600) and P10-F2.5 at 0 T ($Bo_m$ =0). From the magnetorheology studies, we have observed that the shear viscosities of these two fluids are within 3% of one another (for shear range of 0 – 300 s$^{-1}$). Next we focus on the impact of these two fluids at *We* ~100 (fig. 6). Despite the similar viscosities, the P5-F2.5 exhibits rebound suppression at $Bo_m$ ~600, whereas the P10-F2.5 exhibit the typical rebound of SH surface at $Bo_m$ ~0. This clearly illustrates that the rebound suppression is not triggered by the increase in the viscosity of the ferrofluids due to the magnetorheological effect., There is some other non-trivial mechanism at play.

For the explanation, we appeal to our previous report on rebound suppression of elastic fluid droplets on SH surfaces and the references within [12]. We showed that using very dilute polyacrylamide (PAAM) solution droplets, it was possible to induce elastic instability during the retraction phase after impact, which led to rebound suppression. During the spreading phase of the droplet, the long polymer chains unwind due to shear at the spreading contact line, and similar observations have been reported in literature [11]. During the retraction phase, the polymer chains recoil, and the recoiling dynamics is governed by the relaxation timescale of the polymer molecules. If the concentration of the polymer in the solution is such that the relaxation timescale of the non-Newtonian fluid is smaller than the timescale of retraction of the contact line, then the retracting contact line is not subjected to the normal stress generated against the retraction. In such cases, the droplet rebounds as typical on SH surfaces. In the event the polymer concentration is above the threshold, such that the relaxation timescale is greater than the retraction timescale, the triple line retracts faster than the polymer chains, leading to a normal stress generation which decelerates the retraction velocity. In absence of high retraction velocity, the rebound is suppressed. We have also shown that for all such cases of rebound suppression, the associated Weissenberg number ($Wi = \lambda \dot{\gamma}$, where $\lambda$ is the relaxation time of the fluid, and $\dot{\gamma}$ is the shear rate at onset of retraction) is always greater than one. This signifies that the event is triggered by elastic instability within the fluid [12]. As the impact *We* increases, the retraction shear rate increases, and the droplets show higher propensity of rebound arrest. Needless to say, there may be an upper bound to the Weber number beyond which the drops may fragment upon impact on the ground. The present experimental study is performed at Weber number well below this upper bound.

At this juncture, it is noteworthy that such elastic instability can only be triggered in case of very long chain polymer molecules. In the present study, we have used PEG-400, which is a very short chain polymer, and thus the droplets cannot exhibit rebound suppression (via the route of elastic instability) for the range of impact *We* explored. We further confirm this using theoretical analysis and control experiments on the polymer solutions and non-Newtonian ferrofluids in absence of field. In the previous report [12],we have noted that for all cases of rebound suppression, *Wi* >1 is satisfied signifying the onset of elastic instability. Following the same methodology, we deduce the approximate shear rates at the termination of spreading regime and onset of retraction regime from image processing. To determine the relaxation timescales for the polymer solutions, we take the aid of oscillatory rheometry. We first determine the viscoelastic signatures of the polymer solutions and non-Newtonian ferrofluids, and obtain the elastic ($G'$) and viscous ($G''$) moduli of the fluids as function of oscillatory frequency ($\omega$). From the viscoelastic response, we obtain the approximate relaxation times for the different fluids based on established theories [23–25]. While employing the theoretical framework to deduce the relaxation timescales for the non-Newtonian ferrofluids, we have assumed that the nanoparticle based polymer solutions also conform to a similar viscoelastic model as the polymer solutions.

Based on the shear rates and relaxation timescales, we determine the *Wi* and observe that all the values (for different impact *We*) are well below one. This signifies that the elastic instability is absent in the droplets of only polymer solution and the non-Newtonian ferrofluids in absence of field. We have already ruled out the role of increased viscosity under magnetic field as a possible governing agent. Further, we have also noted that there is no rebound suppression in absence of field. Thereby, all evidences lead to the inference that a conjugal effect between the elastic and the magnetic forces is occurring, which leads to the rebound suppression. Next, we apply the same methodology to the different cases of non-Newtonian ferrofluid droplets impacting under field effect. To determine the relaxation timescales under the effect of one particular field strength, we perform frequency sweep oscillatory magnetorheology experiments at different field strengths (fig. S3, supporting information). We determine the values of the *Wi* for different impact velocity and magnetic field strength cases (here we use magnetic field strength instead of *Bo$_m$* as the wide range of properties in presence of field does not allow for the use of a single specific *Bo$_m$* value). Although the *Wi* values are greater than the zero-field cases, it is noted that the values of the *Wi* even for the cases of field induced rebound suppression are below one.

As the *Wi* cannot provide a physical picture of the proposed *magneto-elastic* effect, we propose a modified form of the non-dimensional number to incorporate the effect of the magnetic field. Based on the experimental observations, we propose a new non-dimensional variable, which we term as the magnetic Weissenberg number, expressed as $Wi_M = Wi^{1/2} Bo_m^2$. Physically, the number is the ratio of the magneto-elastic forces to the visco-capillary forces within the non-Newtonian fluid. The exponents for *Wi* and *Bo$_m$* are based on detailed experimental data, and we have selected them in a manner such that the onset of magneto-elastic effect induced rebound suppression happens at the value of 1. The values of $Wi_M$ for P10-F2.5 droplets for different impact velocity and different magnetic field strength have been illustrated in fig. 7a. We observe that the magnetic Weissenberg number criterion is able to segregate the regimes of rebound and rebound suppression ($Wi_M$ >1), with respect to both the impact velocity and applied magnetic field strength. The behaviour of the $Wi_M$ with respect to the magnetic field strength for different non-Newtonian ferrofluid droplets impacting the SH surface at 1.5 m/s has been illustrated in fig. 7b. It can be seen that the proposed $Wi_M$ is able to predict the paradigm of rebound suppression for different non-Newtonian ferrofluids. From the innate definition of the $Wi_M$ and the fact that we can consistently predict rebound suppression at $Wi_M>1$, our hypothesis on the presence of magneto-elastic effect in non-Newtonian ferrofluid impact dynamics is further cemented.

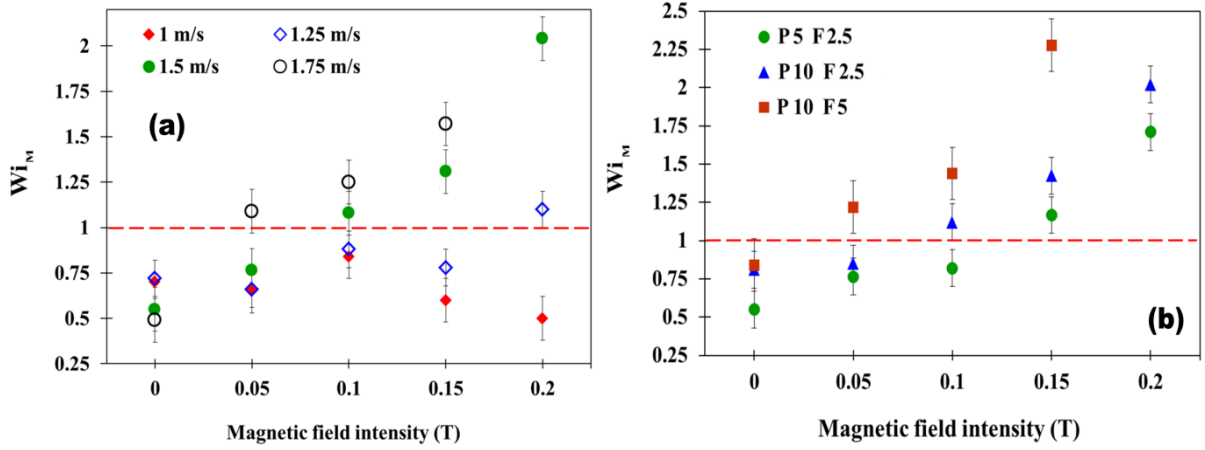

**Fig. 7:** Plot of the magnetic Weissenberg number ($Wi_M$) against the magnetic field intensity, for (a) P10–F2.5 droplets impacting at different velocities, and (b) for different non-Newtonian ferrofluid droplets impacting at velocity of 1.5 m/s. The horizontal dashed line indicated $Wi_M$ =1. All points lying in the regime above this line exhibit rebound suppression triggered by the magneto-elastic effect.

### 3.3. Droplet ferrohydrodynamics regimes

Finally, we elucidate various impact ferrohydrodynamic regimes of non-Newtonian ferrofluid droplets in presence of external uniform horizontal magnetic field. The phase maps for *We* vs $Bo_m$ and $Ca_m$ and *Ha* for different ferrofluids have been illustrated in fig. 8. We discuss the various regimes in the phase map as follows:

***Regime-0:*** In this regime, the needle and droplet assembly lies within the direct influence of magnetic poles. The growing droplet is pulled away from the needle by the magnetic field and distorted largely before the impact, which does not lead to physically consistent observations. Thereby, experiments have not been done in this regime.

***Regime-I:*** In this regime, the droplet rebounds with pinch-off (releasing a tiny droplet from the rebounding parent droplet). This rebound with pinch-off behavior exists due to the dominance of the inertial and capillary forces compared to the magnetic force. Accordingly, this regime is noted to occur at low $Bo_m$ where the magnetic forces are weak, and at low to moderate *We* such that the impact process does not lead to shattering of the droplet due to high inertia.

***Regime-II:*** Complete rebound takes place in this regime, due to the dominance of recoiling kinetic energy of the droplet over viscous dissipation during spreading and the magnetic body force. But in this regime, the capillary force is overshadowed by the rebound inertia, and pinch-off is absent. Consequently, the regime appears in close conjunction with R-I, but extends to relatively higher *We*.

***Regime-III:*** The droplet breaks up during the retraction stage in this regime. In this regime, the droplet spreads to its maximum spread state, and forms radial filaments, which do not detach immediately as the capillary forces are dominated by the viscoelasticity of the fluid. At the maximum spread state, the magnetic force on the bulbs at the ends of the filaments is high at even moderate $Bo_m$. During the retraction stage, the recoil inertia is overcome by the magnetic force, and the bulbs detach off the retracting parent droplet to form smaller droplets. This regime occurs at higher *We* and moderate $Bo_m$ as higher inertia ensures maximum spread state, and the moderate $Bo_m$ ensures the detachment of the filamentous droplets at retraction.

***Regime-IV:*** Fragmentation of the droplet occurs in this regime. This occurs at either high *We* or high $Bo_m$. At high *We*, the impact inertia is high enough to induce fragmentation of the droplet during the spreading state, caused by formation of large velocity gradients within the spreading droplet, which overcomes the capillary and viscous forces. At high $Bo_m$, the magnetic force on the spreading droplet is high enough to induce ferrohydrodynamic instability [8], which leads to rupturing of the spreading droplet as the magnetic forces overcome the capillary forces.

***Regime-V:*** In this regime, suppression of droplet rebound takes place due to the magneto-elastic effect. This occurs at the junctions of R-III and R-IV, where the *We* is moderate enough to induce the elastic instability during the retraction process, and the $Bo_m$ is moderate enough to induce the optimum ferrohydrodynamic force on the droplet. The extent of the regime increases in size with increase in the elasticity of the fluid (polymer concentration) and magnetic moment of the droplet (magnetic particle concentration).

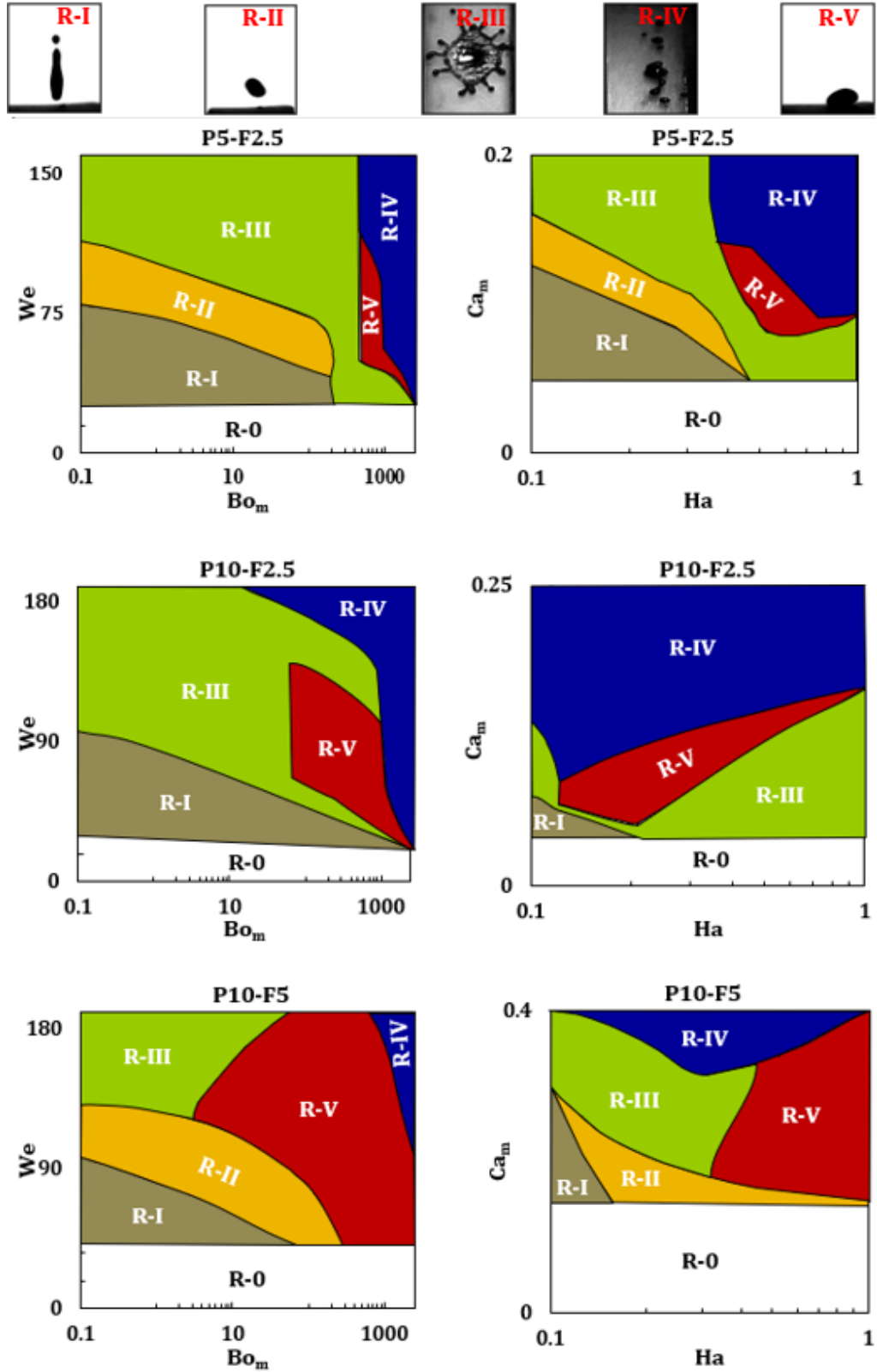

**Fig.8:** Phase maps indicating the different impact ferrohydrodynamic outcomes of non-Newtonian ferrofluid droplets in magnetic field ambience. The regimes R-0, R-I, R-II, R-III, R-IV and R-V represent "non-experimental regime", "rebound with pinch-off", "complete rebound", "breakup during retraction", "fragmentation" (with orthogonal elongation) and "rebound suppression", respectively.

## 4. Conclusions

In this study, we report an extensive experimental investigation on the impact ferrohydrodynamics of non-Newtonian ferrofluid droplets on a SH surface in the presence of a horizontal uniform magnetic field. We used stable colloidal solutions of magnetic nanoparticles dispersed in polymeric solutions as the non-Newtonian ferrofluids. The studies encompass a range of impact *We* up to *~180* and *Bo$_m$* *~0* to *1500*. We noted that in the presence of magnetic field, the non-dimensional maximum spreading ($\xi_{max}$) increases compared to the Newtonian ferrofluids with increasing both polymer and magnetic particles concentration. Through experimental investigations, we have studied the effects of polymer concentration, magnetic nanoparticles concentration, Weber number and *Bo$_m$* on the impact dynamics of non-Newtonian ferrofluid droplet. Addition of polymers to the base Newtonian ferrofluid caused rebound suppression at lower *Bo$_m$* for fixed magnetic particle concentration and *We*. Similarly for fixed polymer concentration and We, increase of magnetic particle concentration triggered rebound suppression at lower *Bo$_m$*.

The combined effect of magnetic particles and elastic effects of polymer chains, together clubbed as magneto-elastic effect, similar to the coinage of elasto-inertial effect of earlier studies [12,26] is shown to be responsible for the early onset of rebound suppression. We formed a non-dimensional number termed as magnetic Weissenberg number, $Wi_M$ taking into account of the effect of classical Weissenberg number and the magnetic Bond number. Through the scaling analysis we showed that when $Wi_M \geq 1$ magneto-elastic instability is triggered and droplet rebound suppression is observed for the first time. This is analogous to the situations when purely elastic instability sets in [27, 28] or onset of drag reduction [29] for Wi≥1. Finally, ferrohydrodynamics of non-Newtonian ferrofluid droplet behavior regime phase maps over a wide range of corresponding dimensionless numbers such as *We*, *Bo$_m$*, *Ca$_m$* and *Ha* were presented. The present findings may have significant implications towards design and development of micro or macroscale systems and devices involving magnetic liquid droplets.

## Data availability statement

All data pertaining to this research are provided in the paper and the supporting information document.

## Acknowledgments
GVVSVP would like to thank the Ministry of Education, Govt. of India, for the doctoral scholarship. DS and PD would like to thank IIT Ropar for funding the present work (vide grants 9-246/2016/IITRPR/144 & IITRPR/Research/193, respectively). PD also thanks IIT Kharagpur for partially funding the work (project code SFI).

## Conflicts of interests
The authors do not have any conflicts of interest with any individuals or agencies for the current research work.

# Supporting information

This document contains additional details on the physical properties of the ferrofluids, additional data on experimental setup, data on their rheology and viscoelastic behavior, etc.

# Supplementary information

**Table. S1:** Surface tension values of the non-Newtonian ferrofluid droplets

| Base fluid | Fe$_3$O$_4$ particles | Surface tension (mN/m) |
|---|---|---|
| **P0** | F2.5 | 72.6 |
|  | F5 | 70 |
| **P5** | F2.5 | 79 |
|  | F5 | 74 |
| **P10** | F2.5 | 69 |
|  | F5 | 69.5 |

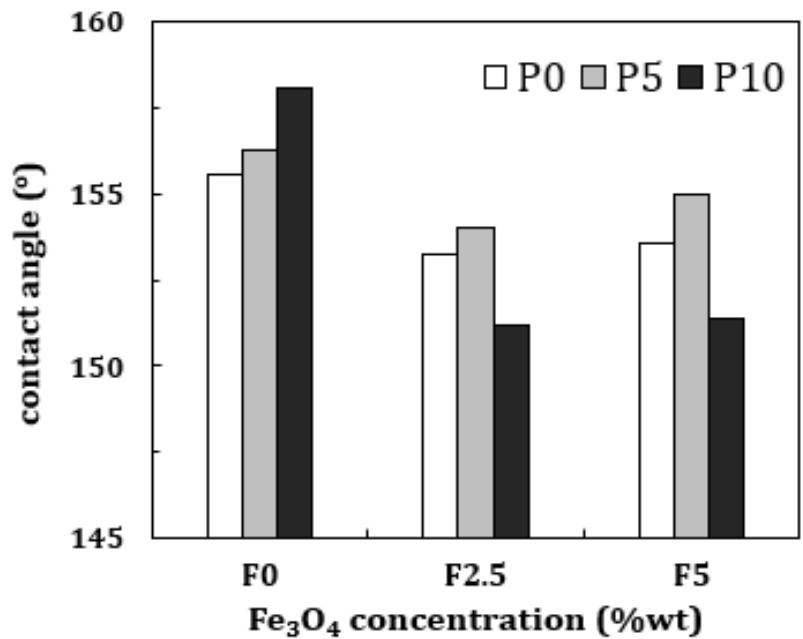

**Fig. S1:** Static equilibrium contact angles of different non-Newtonian ferrofluid droplets on spray coated superhydrophobic surface.

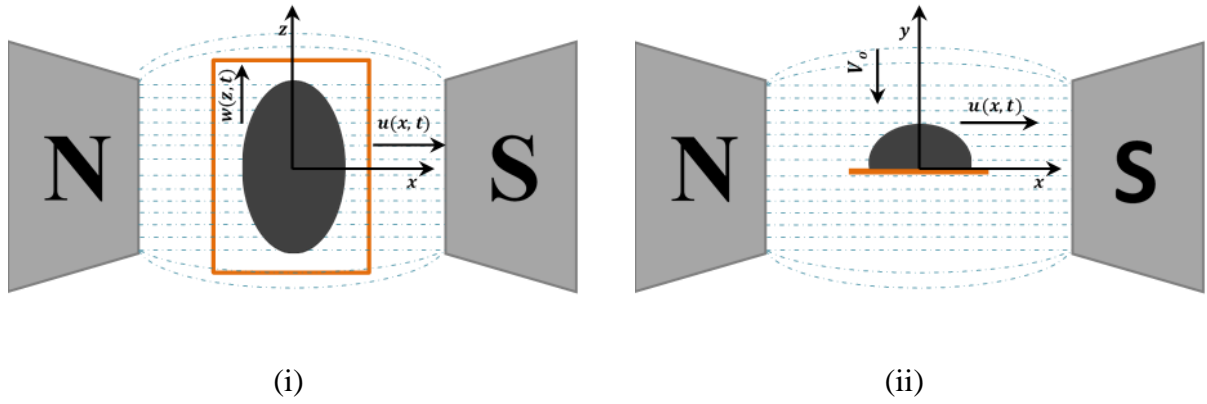

(i)                                                                                  (ii)

**Fig. S2:** (i) top view and (ii) front view of the ferrofluid droplet under magnetic field. N and S represent the north and south poles of the electromagnet.

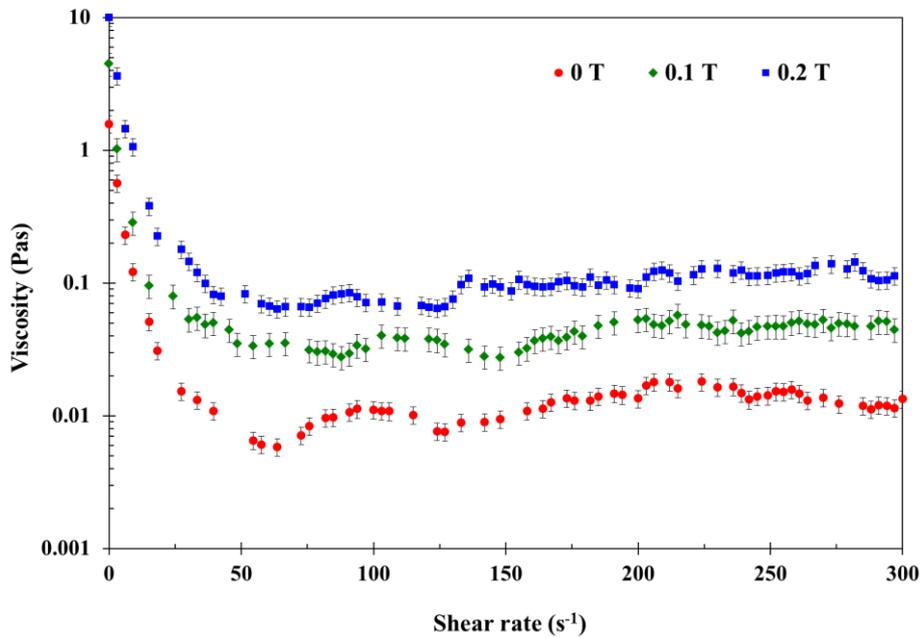

**Fig S3:** Rheological response of the non-Newtonian ferrofluid (P10-F5) in the presence of magnetic field. The sample with the highest polymer and $Fe_3O_4$ content has been chosen to explain the rheology in the context of the most viscous sample. It is noted that the non-Newtonian character of the ferrofluids is retained only up to shear rates of ~50–100 $s^{-1}$. This behaviour is also conserved in case of the magnetorheology. Beyond this regime, the fluids show nearly Newtonian behaviour, albeit with enhanced Newtonian viscosity due to the magnetic field.

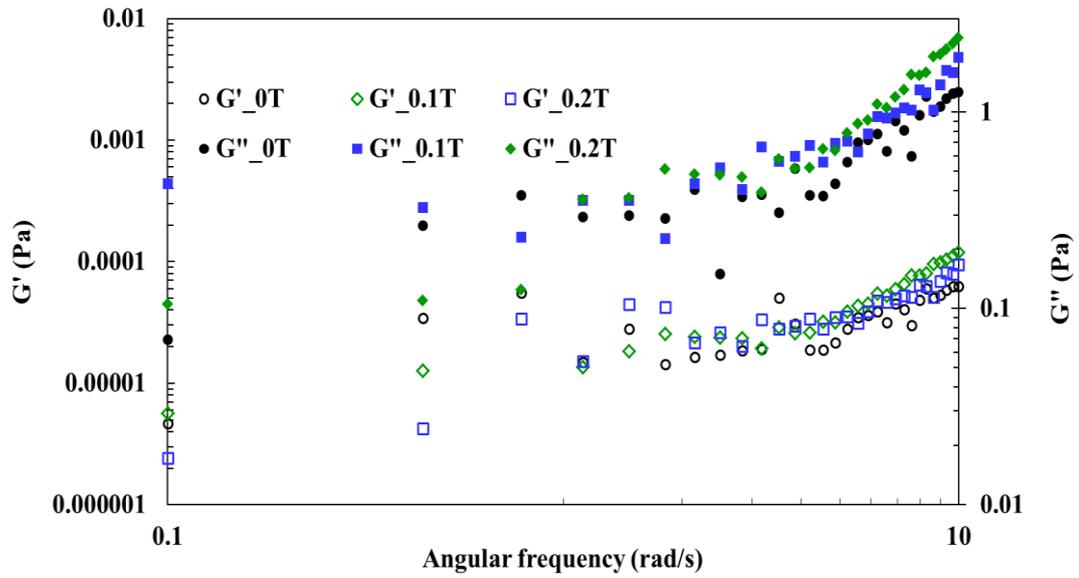

**Fig S4:** Viscoelastic response of the non-Newtonian ferrofluid (P10–F2.5) due to increasing oscillatory angular frequency at 1 % oscillatory strain amplitude. A single case has been illustrated for representation purpose. The oscillatory frequency (ω) dependent values of G' and G" are used to determine the frequency dependent complex viscosity as $\eta_c = \omega^{-1}\sqrt{G'^2 + G''^2}$, which is further used to estimate the relaxation time scales for the non-Newtonian fluids.